\documentclass[a4paper,twocolumn]{article}

\usepackage{cite}             		
\usepackage{url}			

\usepackage{graphicx}
\setkeys{Gin}{width=0.9\columnwidth}	

\begin{document}

\title{Using the Sound Card as a Timer}

\author{C.~E.~Aguiar$^1$ and M.~M.~Pereira$^{1,2}$ \\
$^1$ Instituto de F\'\i sica, Universidade Federal do Rio de Janeiro, Rio de Janeiro, Brasil \\
$^2$ Centro Federal de Educa\c c\~ao Tecnol\'ogica (CEFET-RJ), Nova Igua\c cu, Brasil}

\date{}

\maketitle	

\begin{abstract}
Experiments in mechanics can often be timed by the sounds they produce. In such cases, digital audio recordings provide a simple way of measuring time intervals with an accuracy comparable to that of photogate timers. We illustrate this with an experiment in the physics of sports: to measure the speed of a hard-kicked soccer ball.
\end{abstract}

\section*{Introduction}

Experiments in mechanics often involve measuring time intervals much smaller than one second, a task that is hard to perform with handheld stopwatches. This is one of the reasons why photogate timers are so popular in school labs. There is an interesting alternative to stopwatches and photogates, easily available if one has access to a personal computer with sound recording capability. The idea is simple: a computer sound card records audio frequencies up to several kilohertz, which means it has a time resolution of a fraction of millisecond, comparable to that of photogate timers. Many experiments in mechanics can be timed by the sound they produce, and in these situations a direct audio recording may provide accurate measurements of the time intervals of interest. This idea has already been explored in a few cases \cite{Stensgaard,Cavalcante,Aguiar1,Hunt,White,Barrio}, and here we apply it to an experiment that our students found very enjoyable: to measure the speed of soccer balls kicked by them.

\section*{How hard can you kick the ball?}

The physics of sports never fails to catch the attention of students (at least for a while). In particular, measurements of their performance in a game can attract a lot of interest. Here we describe a method of measuring how fast they kick a soccer ball. We are interested in the speed of the ball immediately after it is hit, not on the long term average speed. The later depends on many factors—bounces on the ground, mean height, air drag—that make it difficult to compare different shots. The average speed will not depend so much on these details if we measure it over a short distance from the kick point. In this case the result is a good estimate of the speed the ball gains when hit. But there is a problem now: it is difficult to measure the time it takes for a hard-kicked ball to travel a short distance (2--3 m). Typical times are of the order of a tenth of a second, so that handheld stopwatches cannot be used efficiently. Even photogate timers are not very useful here---they probably demand some reshaping in order to accommodate for the size of the ball and its potentially destructive effects. 

\begin{figure}[ht]
\begin{center}
\includegraphics[width=0.65\columnwidth]{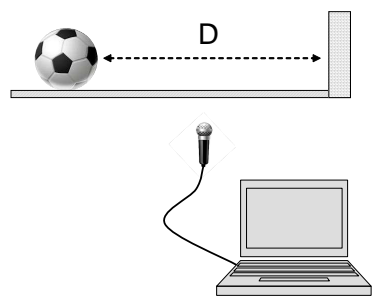}
\caption{Setup for measuring the ball speed.}
\label{setup} 
\end{center}
\end{figure}

\begin{figure}[thb]
\begin{center}
\includegraphics[width=0.75\columnwidth]{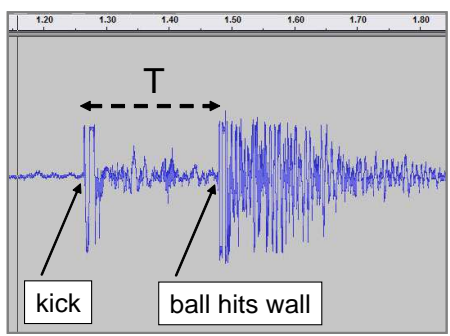}
\caption{Waveform recorded in a measurement of the ball speed. The two sound pulses are produced by the foot kicking the ball and by the ball hitting the wall. The time scale at the top is in seconds.}
\label{waveform} 
\end{center}
\end{figure}

The computer sound card provides a simple way to perform the measurement. The setup is shown in Fig.~\ref{setup}. The ball is placed at a known distance $D$ from a wall, and the computer (with a microphone plugged to the sound card) is set on a safe spot roughly equidistant from the ball and the wall. The measurement consists basically on having a student kick the ball (aiming at the wall) while the computer records the sound input. There are many programs that can be used not only to control the recording but also to display and analyze the resulting audio data (the sound editor \emph{Audacity} \cite{Audacity} is a good and free choice). Figure~\ref{waveform} shows the waveform recorded during a shot. Two distinct sound pulses are seen (and heard), corresponding to the foot hitting the ball and the ball hitting the wall. The start time of each pulse can be accurately determined---to a fraction of millisecond---by zooming in on the corresponding sector of the waveform (all sound editors have zoom tools). The time-of-flight of the ball, T, is the difference between these two times, as shown in Fig.~\ref{waveform}. Given the time-of-flight $T$ and distance $D$, the speed of the ball is $V = D/T$.  For the shot seen in Fig.~\ref{waveform}, $D = 2.5$~m and $T = 0.214$~s, so that $V = 42$~km/h.

\begin{figure}[thb]
\begin{center}
\includegraphics{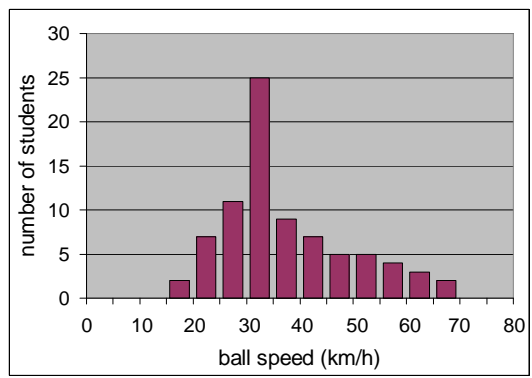}
\caption{Speed distribution of the kicked balls. The sample corresponds to 80 high school students, each taking one shot.}
\label{histogram} 
\end{center}
\end{figure}

\begin{figure}[thb]
\begin{center}
\includegraphics{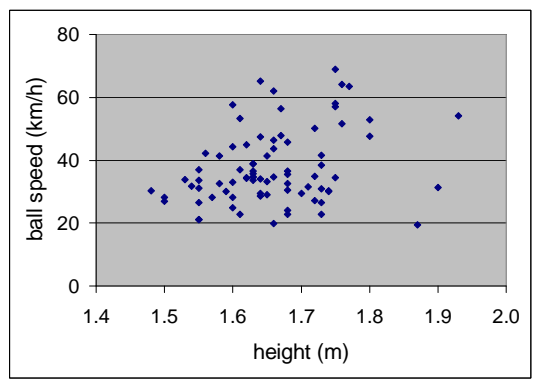}
\caption{Ball speed vs. height of the student who kicked it.}
\label{scatterplot} 
\end{center}
\end{figure}

It is interesting to have the students perform statistical analyses of the data they collect. The histogram in Fig.~\ref{histogram} shows the distribution of ball speeds in a group of 80 students (from three different classes of a high school). The mean speed of the balls kicked by them was 38 km/h, with a standard deviation of 12 km/h.  The students were encouraged to look for correlations between their shooting performances and personal characteristics like age, gender, height or mass. Somewhat surprisingly, only very weak (if any) correlations were found in this sample. The scatter plot in Fig.~\ref{scatterplot} shows the ball speed vs. the height of the student who kicked it. Simple inspection of the plot suggests that the height of a student doesn't have much influence on how hard he kicks the ball. A more refined analysis confirms this---the correlation coefficient of the data is $r^2 = 0.09$. 

\section*{Comments}

The measurement was quite easy to perform even with relatively large groups of students. We used a number of small laptops (netbooks), which allowed teams of 3--4 students to share a computer for data taking and analysis. As the experiment is better executed in a gymnasium or in open space, portable computers are a good choice whenever possible. If these are not available (or not in sufficient number), one can do the sound recording with an \emph{mp3} recorder/player---a gadget that students frequently carry with them---and later transfer the audio files to a computer for analysis.

The time-of-flight measurement with the sound card is very accurate, so that the error in the calculated speed comes mostly from uncertainties in the distance traveled by the ball. For most kicks the ball doesn't head straight to the wall, and the length $D$ shown in Fig.~\ref{setup} is really a lower bound to the traveled distance. To minimize this uncertainty, it is useful to draw a target area on the wall and accept only shots that hit this region. If the target area is sufficiently small, it is possible to keep the distance uncertainty within acceptable limits. 

Measurements of the ball speed are a good starting point for many interesting discussions on the physics of soccer kicks. One could, for example, investigate if air drag is important at the typical speeds found in the experiment, or try to estimate how fast the foot must move in order to impart such speeds to the ball. Several topics that can be explored in this manner are found in Refs.~\cite{Wesson,Ireson,Aguiar2,Vieira}. 

\section*{Acknowledgments}
We thank Funda\c c\~ao de Amparo \`a Pesquisa do Rio de Janeiro (Faperj) for financial support.

\end{document}